\documentclass{amsart}
\def\dblsp{\baselineskip2\baselineskip\lineskip2\lineskip}
\let\dblsp\relax
\def\C{\mathbb{C}}
\def\bibref[#1]{\cite{#1}}

\newtheorem{theorem}{Theorem}[section]

\makeatletter
\def\@oddhead{\underline{\hbox to \textwidth{On the Quantization of a Self-Dual
Integrable System\hfill Kasman}}}
\let\@evenhead\@oddhead

\def\@oddfoot{$\overline{\hbox to \textwidth{\hfill
\bf\thepage\hfill}}$}
\let\@evenfoot\@oddfoot
\advance\footskip.5in
\makeatother

\begin{document}

\dblsp

\title{On the Quantization of a Self-Dual\\ Integrable System\\ \ \\
\tiny (to appear \textit{Journal of Physics A}, 2001)}

\author{Alex Kasman}

\address{Department of Mathematics\\College of Charleston}

\begin{abstract}\normalsize\dblsp

In this note, we apply canonical quantization to the self-dual particle system describing the motion of poles to a higher rank solution of the KP hierarchy, explicitly
determining both the quantum Hamiltonian and the wave function.  It is
verified that the quantum Hamiltonian is \textit{trivially bispectral}
(that is, that the wave function can be taken to be symmetric) as
predicted by a widely held hypothesis of mathematical physics.

\end{abstract}
\maketitle

\section{Introduction}

Following Ruijsenaars \bibref[Ruij], it has been recognized as
convenient and useful to classify integrable particle systems into
\textit{dual pairs}.  Roughly speaking, one system is dual to another
when the linearizing map of one system is the inverse of the
linearizing map of the other.  More specifically, a system is said to
be \textit{self-dual} if it is linearized by an involution.  (The
Calogero-Moser particle system is the best known self-dual integrable
system \bibref[AMcM].)  Duality of integrable systems has been the
focus of much research lately due to its role in theories of quantum
gravity (see, for example, \bibref[B3M,Marsh]).  In particular, one observation is that the
quantized version of these systems should demonstrate a symmetry of
spatial and spectral parameters in their wave functions
\bibref[Gorsky-Rubtsov-Fock-Nekrasov,HK].  Since this is a special
instance of the \textit{bispectral property} \bibref[BispBook], this
conjecture will be referred to here as the \textit{bispectral
quantization hypothesis} (BQH).

Although there is no particular reason to disbelieve BQH, it is
nowhere supported by a mathematical proof.  (In fact, the ambiguities
of the procedure known as `quantization' would make it difficult to
state the BQH in a verifiable manner.)  At present, at
least, it is supported only `experimentally' by the fact that it
happens in all the well-known systems (Calogero-Moser,
Ruijsenaars-Schneider, Sutherland, etc.) and by the fact that it
`seems like the appropriate analogue'.  It is therefore of interest,
when confronted with a new example of a pair of integrable systems
related by classical duality, to determine whether it might provide a
counter-example.  The purpose of this paper is to confirm that the quantum
Hamiltonian
\begin{equation}
\tilde{\mathcal{H}}=\left(\sum_{i=1}^n\partial_i^2-x_i\right)-\left(\sum_{1\leq i<j\leq
n}\frac{4}{(x_i-x_j)^2}\right)\label{eqn:tilcalH}
\end{equation}
(corresponding to a self-dual classical system first explicitly
studied in \bibref[R]) does have a wave function with the predicted
symmetry.  
In addition, this result is of interest since it provides
new examples of bispectral commutative rings of partial differential
operators containing Schr\"odinger operators and demonstrates that the
operator $\tilde{\mathcal{H}}$ intertwines with the operator
${\mathcal{H}}=\sum_{i=1}^n\partial_i^2-x_i$ (in contrast to the case
$n=1$ where ${\mathcal{H}}$ is known not to intertwine with any other
rational operators.)

\section{Classical Particle Dynamics}\label{sec:classical}

In this section of the paper we will be concerned with the
$n$-particle dynamical system determined by the Hamiltonian:
$$
{H}=\left(\sum_{i=1}^ny_i^2-x_i\right)-\left(\sum_{1\leq i<j\leq
n}\frac{4}{(x_i-x_j)^2}\right)
$$
where $x_i$ are the particle positions and $y_i$ are their momenta.
This Hamiltonian bears a clear resemblence to the Calogero-Moser
particle system \bibref[AMcM,cmbis,Kr,Shiota,W2].  In fact, it is
similar to that famous system in several important ways.

Most importantly, we will observe that this system -- like
Calogero-Moser -- is integrable and self-dual.  To see this, it is
convenient to write $H$ in terms of the Calogero-Moser Matrices
\bibref[KKS,W2].  That is, consider the set of all pairs of $n\times n$ matrices $(X,Z)$
satisfying
\begin{equation}
\textup{rank}([X,Z]+I)=1.\label{eqn:rk1}
\end{equation}
An element of this set can naturally be associated to a state of the system in
which the particles occupy distinct positions.  In that case we consider:
\begin{equation}
X=x_i \delta_{ij}\qquad
Z= y_i\delta_{ij}+\frac{\sqrt{2}(1-\delta_{ij})}{x_i-x_j}\label{eqn:XZ}
\end{equation}
(Note that as compared to the presentation of these matrices in other
papers, a factor of $\sqrt{2}$ has been added to the matrix $Z$ off of
the diagonal for later convenience.)

It was observed in \bibref[R] that as in the case of
Calogero-Moser, the Hamiltonian function can be
written simply in terms of $X$ and $Z$:
$$
{H}=\textup{Tr}(Z^2-X).
$$
Now observe that the map
$$ (X,Z)\mapsto(\bar X,\bar Z)=((Z^{\top})^2-X^{\top},Z^{\top}) $$
is an involution on the space of matrices satisfying the rank one
condition \eqref{eqn:rk1}.  More importantly, note that the
corresponding Hamiltonian is
$$ \bar H_j=\textup{tr}((\bar Z)^2-\bar X)^j =\textup{tr}(X^j)=\sum
x_i^j. $$ Since the Hamiltonian is independent of the $y$'s, the $x$'s
are constant while the $y$'s change linearly.  Since this map is a
linearizing involution, we are thus able to
conclude that ${H}$ is a self-dual integrable system.

Another way in which this system is similar to Calogero-Moser is that
both systems describe the motion of poles in a solution to the KP
hierarchy.  In particular, as shown in \bibref[R], this Hamiltonian
describes the dynamics of poles to a solution determined as an
iterated Darboux transformation of the Airy solution \bibref[KR].
Unlike Calogero-Moser, however, this solution is not a rational
function of all of the time variables of the KP hierarchy \textit{and}
this solution does not correspond to a flow on the Jacobian variety of
the spectral curve.  In fact, the corresponding KP solution is associated to
rank \textit{two} bundles over the spectral curve, rather than rank
one bundles as in the case of all particle systems for which the BQH
has previously been tested.

\subsection{Bispectrality and Duality}\label{sec:duality}

A linear differential operator $L$ in the variables $\vec
x=\{x_1,\ldots,x_n\}$ is said to be \textit{trivially bispectral}
if there is a non-zero family of eigenfunctions $\psi(\vec x,\vec
z)$ 
parameterized by $\vec z=\{z_1,\ldots,z_n\}$ such that
$$
L\psi(\vec x,\vec z)=p(\vec z)\psi(\vec x,\vec z)\qquad \textup{and}
\qquad
\psi(\vec x,\vec z)=\psi(\vec z,\vec x).
$$
This is a special case of the more general bispectral property first
considered in \bibref[DG] which does not require that $\psi$ be
symmetric, but only that it should also be an eigenfunction for an
operator $\Lambda$ in $\vec z$ with eigenvalue depending on $\vec x$.
(See \bibref[BispBook] for a recent overview of this
field and its diverse connections to mathematical physics.)

Bispectrality is related to the duality (cf.\ \bibref[HK,Ruij]) of
integrable particle systems both at the classical and the quantum
levels.  The main result of this paper concerns the quantum
manifestation, but its role in the classical case is also relevant as
motivation for the main result.  Therefore, let us recall that the
self-duality of the classical Calogero-Moser system was related to
bispectrality in \bibref[cmbis] and \bibref[W2] where the linearizing
map for this system was found to be equivalent to the exchange of
spectral and spatial parameters in the KP wave function.  In contrast,
the self-duality of the higher rank classical systems presented above
was conjectured in \bibref[KR] due to the fact that they too describe
the motion of the poles of bispectral KP Lax operators.

It is interesting that the bispectral problem led us to self-dual
classical integrable systems because bispectrality is really the
structure of \textit{quantum} duality.  That is, according to the BQH
\bibref[Gorsky-Rubtsov-Fock-Nekrasov], dual systems when quantized
should share an eigenfunction with spatial and spectral parameters
reversed (and hence self-duality should be manifested as trivial
bispectrality).  A naive form of canonical quantization of the
Hamiltonian function ${H}$ involves formally replacing $y_i$
with the differential operator $\partial_i$.  This then leads us to
the main question:

\textit{
Is there a non-zero eigenfunction $\tilde\psi$ satisfying the wave equation
$$
\tilde {\mathcal{H}}\tilde\psi(\vec x,\vec z)=p(\vec z)\tilde\psi(\vec x,\vec z)
$$
for some non-constant function $p$
and the operator $\tilde{\mathcal{H}}$ \eqref{eqn:tilcalH} 
with the symmetry
\begin{equation}
\tilde\psi(\vec x,\vec z)=\tilde\psi(\vec z,\vec x)\hbox{?}\label{eqn:symmetry}
\end{equation}}

\section{Intertwining Relations}\label{sec:intertwining}

The operators $L$ and $\tilde L$ are said to be \relax{intertwined
by} $K$ if they satisfy $$ KL=\tilde LK.  $$ Such relationships are
useful as \textit{transformations} for producing an operator $\tilde L$
with specified spectral properties from a known $L$.  For instance,
this is one way to derive the quantum Calogero-Moser operator (the
canonical quantization of the \textit{second} Calogero-Moser
Hamiltonian function) \bibref[CV].

One begins with the constant coefficient operator $$
\Delta=\sum_{i=1}^n \partial_i^2\in \C[\partial_1,\ldots,\partial_n]
$$ in the commutative ring of constant coefficient operators.  Every
element of this ring has the function $$ \phi(\vec x,\vec
z)=\exp(x_1z_1+\cdots+x_nz_n) $$ as an eigenfunction with eigenvalues
depending on $\vec z$ and this is perhaps the most elementary example
of trivial bispectrality.
The following theorem of Chalykh and Veselov is provided here not only
as an example but also as an important lemma.

\begin{theorem}\label{thm:CV}{\bibref[CV] There is a partial differential operator
$\mathcal{D}_n$ ($n\geq2$) which takes the form of a polynomial in
$\partial_{ij}=\partial_i-\partial_j$  with
coefficients rational in $x_{ij}=x_i-x_j$ ($1\leq i<j\leq n$) such
that
$$
\mathcal{D}_n\Delta=\tilde\Delta\mathcal{D}_n\qquad
\tilde\Delta=\Delta-\sum_{1\leq i<j\leq
n}4x_{ij}^{-2}.
$$
Then the function
$$
\tilde\phi(\vec x,\vec z)=\prod_{1\leq i<j\leq n} (z_i-z_j)^{-1}
\mathcal{D}_n[\phi(\vec x,\vec z)]
$$
is an eigenfunction for the Calogero-Moser hamiltonian operator
$\tilde \Delta$ satisfying $\tilde\phi(\vec x,\vec z)=\tilde
\phi(\vec z,\vec x)$.
}\end{theorem}

It is an immediate consequence of that theorem that
one can similarly construct  the quantized
Hamiltonian $\tilde {\mathcal{H}}$ via an intertwining relationship:
\begin{theorem}\label{thm:intertwines}{Let
$$
{\mathcal{H}}=\sum_{i=1}^n\partial_i^2-x_i
\qquad\hbox{and}\qquad
\tilde {\mathcal{H}}= {\mathcal{H}}-\sum_{1\leq i<j\leq n} 4(x_i-x_j)^{-2}.
$$
Then the operator $\mathcal{D}_n$ from Theorem~\ref{thm:CV} satisfies
$$
\mathcal{D}_n {\mathcal{H}}=\tilde {\mathcal{H}}\mathcal{D}_n.
$$}
\end{theorem}
\begin{proof}
Since $[\partial_i-\partial_j,x_1+x_2+\cdots+x_n]=0$, one has that
$$
\mathcal{D}_n{\mathcal{H}}=
\mathcal{D}_n(\Delta-(x_1+\cdots+x_n))
=(\tilde \Delta-(x_1+\cdots+x_n))\mathcal{D}_n
=\tilde{\mathcal{H}}\mathcal{D}_n.$$
\end{proof}

\noindent\textbf{Remark:} This is interesting to contrast with the one-dimensional
case.  The ordinary differential operator $\partial^2$ and
$\partial^2-2/x^2$ are intertwined by the operator $\partial-1/x$,
which can be regarded as the one dimensional case of
Theorem~\ref{thm:CV}.  In contrast,  the Airy
operator $\partial^2-x$ is unique among bispectral Schr\"odinger
operators (cf. \bibref[DG]) in that it
\textit{cannot} be intertwined with another rational operator.
Therefore, it may be seen as a somewhat surprising fact that ${\mathcal{H}}$,
its higher dimensional analogue, does intertwine with a rational
operator.

\section{Eigenfunctions}\label{sec:eigenfunctions}

Still, this does not resolve the question of whether $\tilde
{\mathcal{H}}$ has an eigenfunction which is symmetric in spatial and
spectral parameters.  Unfortunately, the first thing one might try
turns out to be a `wrong turn'.  Since the operator ${\mathcal{H}}$ is
contained in the commutative ring
$\C[\partial_1^2-x_1,\ldots,\partial_n^2-x_n]$ (polynomials in $n$
different one-dimensional Airy operators) which has the symmetric
common eigenfunction $$ \sigma(\vec x,\vec z)=\prod_{i=1}^n
\textup{Ai}(x_i+z_i), $$ one has immediately that
\begin{theorem}\label{thm:falselead}{The operator $\tilde {\mathcal{H}}$ has
eigenfunction $\tilde \sigma=\mathcal{D}_n[\sigma]$ satisfying the
equation $$ \tilde{\mathcal{H}}\tilde\sigma=(\sum z_i)\tilde\sigma.  $$}
\end{theorem}
 However,
neither $\tilde\sigma$ \textit{nor} any multiple of it by a non-zero
function of $\vec z$ is symmetric in spatial and spectral parameters.
This apparent counter-example to the bispectral quantization
hypothesis is resolved by recognizing that ${\mathcal{H}}$ is also
contained in another commutative ring with another symmetric common
eigenfunction.

\begin{theorem}\label{thm:betterring}{
The operators $\partial_{in}=\partial_i-\partial_n$ ($1\leq i<n$) each
commute with ${\mathcal{H}}$
and the commutative ring
$\C[\partial_{1n},\ldots,\partial_{n-1\,n},{\mathcal{H}}]$ has the
symmetric common eigenfunction
$$
\psi(\vec x,\vec z)
=\exp\left(\frac{1}{n}\sum_{1\leq i<j\leq n} x_{ij}
z_{ij}\right)\textup{Ai}\left((\frac{1}{n})^{1/3}\sum_{i=1}^n
(x_i+z_i)\right)
$$
satisfying
$$
\partial_{ij}\psi=z_{ij}\psi=(z_i-z_j)\psi
\qquad
\textup{and}
\qquad
{\mathcal{H}}\psi=p_n(\vec z)\psi
$$
for the polynomial 
$$
p_n(\vec z)=\sum_{j=1}^n\left( \left(\sum_{i=1}^n
z_{ij}\right)^2+z_j\right).
$$}
\end{theorem}
\begin{proof}
The easiest way to observe this is by direct
computation.  For instance, one may derive $p_n$ by first noting that
$$
(\partial_j^2-x_j)\psi=\left((\sum z_{ij})^2+\frac1n\sum(x_i+z_i)-x_j\right)\psi+(\sum z_{ij})\psi'
$$
for a function $\psi'$ that does not matter since it disappears
when one sums over $j$.  However,a more instructive way to verify the
claim is to consider
the change of variables $\alpha_i=x_i-x_n$ ($1\leq i<n$) and
$\alpha_n=x_1+x_2+\ldots+x_n$  after which the operator ${\mathcal{H}}$ decomposes
into a sum of a constant coefficient operator in $\alpha_i$ for $i<n$
and an Airy operator in $\alpha_n$.
\end{proof}

\begin{theorem}\label{thm:symef}{The 
function $$\tilde\psi(\vec x,\vec z)=\prod_{i<j}z_{ij}^{-1}
\mathcal{D}_n [\psi]$$
satisfies the eigenvalue equation
$\tilde{\mathcal{H}}\tilde\psi=p_n(\vec z)\tilde \psi
$
and is symmetric (satisfying \eqref{eqn:symmetry}).}
\end{theorem}
\begin{proof}
Note that every differential operator $D(\vec x,\vec \partial)$ acting on the function
$\phi=\exp(\sum x_iz_i)$ merely acts as a multiplication operator,
multiplying by the polynomial $D(\vec x,\vec z)$ in $\vec z$ with coefficients that
depend on $\vec x$.  This is not necessarily true for differential
operators acting on $\psi$, but it is true that
 a differential operator which
can be written as a polynomial in $\partial_{ij}$ (as $\mathcal{D}_n$
can by construction) just multiplies $\psi$ by this same polynomial.
In particular, $\tilde\phi/\phi=\tilde\psi/\psi=\mathcal{D}_n(\vec
x,\vec z)$ is the same polynomial in $\vec z$ and so the symmetry of
$\tilde \psi$ is equivalent to the already verified symmetry of
$\tilde\phi$. 
\end{proof}

\subsection{Conclusion}

We have seen that the self-duality of the quantum
Hamiltonian $\tilde{\mathcal{H}}$ is directly related to the
separability of the quantum Calogero-Moser Hamiltonian (specifically,
the fact that $\tilde \Delta$ commutes with any operator in the
variable $\alpha_n=x_1+\cdots+x_n$).
This sheds
light on the relationship between the system ${\mathcal{H}}$
and the  Calogero-Moser system, and provides additional support for bispectral quantization hypothesis.

Note that this $\tilde{\mathcal{H}}$ and $\tilde\Delta$ agree on the
hyperplane $\sum_{i=1}^n x_i=0$ (which Chalykh-Veselov consider as a
hypothesis anyway in order to avoid separability).  Still, there
remain essential differences between the two.  In particular, one
finds differences when considering the algebraic structure of system
since, in contrast to the Calogero-Moser system, there is no first
order operator commuting with $\tilde{\mathcal{H}}$.

The Hamiltonian system quantized above is only one example of
many that were suggested by the results in \bibref[KR].  See
\bibref[R] for a hierarchy of classical Hamiltonian functions (of
arbitrary order) which are linearized by an involution and hence
should quantize bispectrally.  

\bigskip

\noindent\textbf{Acknowledgement:}  I am grateful for advice from and helpful
conversations with H. Braden, J. Harnad, A. Marshakov, and J.L. Miramontes.

\end{document}